\documentstyle[aps,12pt]{revtex}
\begin{document}
\title{Nested Domain Defects}
\author{J.R. Morris}
\address{{\it Physics Department, Indiana University Northwest,}\\
{\it 3400 Broadway, Gary Indiana 46408, USA}\\
\bigskip\ }
\maketitle

\begin{abstract}
An example of a supersymmetric model involving two interacting chiral
superfields is presented here which allows for solutions describing
string-like ``domain ribbon'' defects embedded within a domain wall. It is
energetically favorable for the fermions within the wall to populate the
domain ribbons, and an explicit solution is found for the fermion zero
modes. The Fermi gas within ribbons can allow them to stabilize in the form
of small loops.

\smallskip\ 

PACS: 11.27.+d, 12.60.Jv, 98.80.Cq

\bigskip\ 
\end{abstract}

\section{Introduction}

Much attention has been given to topological defects, not only because they
are interesting nonperturbative solutions in field theories, but also
because they may have been physically realized in the early Universe\cite
{vsbook,ktbook}. The additional possibility that the early Universe may have
existed in a supersymmetric phase provides motivation to investigate
possible types of defects that may occur in supersymmetric theories. Here
attention is focused on domain defects that can arise from broken discrete
symmetries, and in particular, we investigate the case of nested domain
defects wherein a string-like, or ribbon-like, defect referred to here as a
``domain ribbon'' can inhabit the interior of a domain wall. Recently,
domain ribbons have been looked at in a nonsupersymmetric theory\cite
{morrisrib}, and in a model\cite{bazeia1} that can be seen as the real
bosonic sector of a supersymmetric theory. Within this context, the present
investigation serves as an extension of these previous studies.
Supersymmetric theories with a single chiral superfield can admit domain
wall solutions with some interesting properties\cite{cveticprl,cvetic}, and
an inclusion of a second chiral superfield allows nontrivial interactions
that can result in a nontrivial internal structure of a domain wall.
Furthermore, a supersymmetric theory naturally includes fermions which can
interact with the scalar fields in interesting ways.

As an example, we assume a relatively simple superpotential which gives rise
to a model which, in the real bosonic sector, admits domain ribbon
solutions. The domain ribbons appear because the system stabilizes by
forming a real scalar condensate in the wall's core. However, the condensate
formation breaks a discrete $Z_2$ symmetry so that different condensate
domains can form in initially uncorrelated regions of the wall, and these
different domains must be separated by a ``wall within the wall'', i.e. a
domain ribbon. The fermions can respond to the scalar field background by
forming zero modes\cite{jackiw}, for which analytical solutions are
obtained. It becomes energetically favorable for fermions to populate the
ribbons, where they are massless. Consequently, in the supersymmetric model,
a Fermi gas pressure can exist within the ribbons, modifying the eventual
fate of the ribbons. Instead of rapidly fissioning away to nothing, as in a
nonsupersymmetric model\cite{morrisrib}, fermion supported closed ribbons
can stabilize in the form of small loops, which may be particle-sized. A
domain wall may therefore end up being populated with these small Fermi
loops.

A supersymmetric model with two interacting chiral superfields is presented
in the next section, and in sec. III the domain wall and domain ribbon
solutions are found in the real bosonic sector. The fermion zero modes,
which can behave as traveling waves that propagate through the ribbons, are
then analyzed in sec. IV. In sec. V we consider an effective one dimensional
Fermi gas in a ribbon, and look at the conditions for which a closed ribbon
loop can stabilize. We conclude with a brief summary.

\section{A Model with Two Interacting Chiral Superfields}

\subsection{Fields}

We consider a supersymmetric theory involving the two chiral superfields $%
\Phi $ and $X$. These chiral multiplets can be displayed as $\Phi =(\phi
,\psi _\phi ,F_\phi )$, $X=(\chi ,\psi _\chi ,F_\chi )$, where $\phi $ and $%
\chi $ are complex-valued scalar fields, $\psi _{\phi ,\chi }$ are Weyl
2-spinors, and $F_{\phi ,\chi }$ are complex-valued scalar auxiliary fields.
From a superpotential $W(\Phi ,X)$, a scalar potential $V(\phi ,\chi )$ can
be generated describing the interactions between the scalar fields. The
Yukawa couplings for the fermions are obtained from the superpotential $W$.
The two chiral superfields $\Phi $ and $X$ have superspace representations%
\cite{srivastava,wbbook} 
\begin{equation}
\begin{array}{l}
\Phi (y,\theta )=\phi (y)+\sqrt{2}\theta \psi _\phi +\theta ^2F_\phi (y), \\ 
X(y,\theta )=\chi (y)+\sqrt{2}\theta \psi _\chi +\theta ^2F_\chi (y),
\end{array}
\label{e1}
\end{equation}

\noindent where $y^\mu =x^\mu +i\theta \sigma ^\mu \bar{\theta}$. (We use a
metric with signature $(+,-,-,-)$. See the Appendix for notation,
conventions, and gamma matrices.) The complex scalars $F_{\phi ,\chi }$ are
auxiliary fields which will be eliminated. Majorana 4-spinors $\Psi _{\phi
,\chi }$ can be defined in terms of the Weyl 2-spinors: 
\begin{equation}
\Psi _\phi =\left( 
\begin{array}{l}
\psi _{\phi \,\alpha } \\ 
\bar{\psi}_\phi ^{\dot{\alpha}}
\end{array}
\right) ,\,\,\,\,\,\,\,\Psi _\chi =\left( 
\begin{array}{l}
\psi _{\chi \,\alpha } \\ 
\bar{\psi}_\chi ^{\dot{\alpha}}
\end{array}
\right) ,\,\,\,\,\,\,\,\alpha ,\dot{\alpha}=1,2.  \label{e2}
\end{equation}

\subsection{Lagrangian}

In terms of the chiral superfields, the Lagrangian is 
\begin{equation}
L=(\Phi ^{*}\Phi )|_{\theta ^2\bar{\theta}^2}+(X^{*}X)|_{\theta ^2\bar{\theta%
}^2}+W|_{\theta ^2}+W^{*}|_{\theta ^2},  \label{e3}
\end{equation}

\noindent where $W$ is the superpotential and $W|_{\theta ^2}$ represents
the $\theta ^2$ part of $W$, etc. In terms of the component fields, $L$ can
be written as 
\begin{equation}
L=L_K^B+L_K^F+L_Y-V,  \label{e4}
\end{equation}

\noindent where 
\begin{equation}
L_K^B=\partial ^\mu \phi ^{*}\partial _\mu \phi +\partial ^\mu \chi
^{*}\partial _\mu \chi ,  \label{e5}
\end{equation}
\begin{equation}
L_K^F=\frac i2\left[ (\partial _\mu \psi _\phi )\sigma ^\mu \bar{\psi}_\phi
-\psi _\phi \sigma ^\mu \partial _\mu \bar{\psi}_\phi +(\partial _\mu \psi
_\chi )\sigma ^\mu \bar{\psi}_\chi -\psi _\chi \sigma ^\mu \partial _\mu 
\bar{\psi}_\chi \right] ,  \label{e6}
\end{equation}
\begin{equation}
\begin{array}{ll}
L_Y & =-\frac 12\sum\limits_{i,j}\left[ Y_{ij}\psi _i\psi _j+Y_{ij}^{*}\bar{%
\psi}_i\bar{\psi}_j\right] \\ 
& =-\frac 12\left[ Y_{\phi \phi }\psi _\phi \psi _\phi +Y_{\chi \chi }\psi
_\chi \psi _\chi +2Y_{\phi \chi }\psi _\phi \psi _\chi \right] +c.c
\end{array}
,  \label{e7}
\end{equation}
\begin{equation}
V=|F_\phi |^2+|F_\chi |^2=\left| \frac{\partial W}{\partial \phi }\right|
^2+\left| \frac{\partial W}{\partial \chi }\right| ^2,  \label{e8}
\end{equation}

\noindent with $Y_{ij}=\frac{\partial ^2W}{\partial \varphi _i\partial
\varphi _j}$, $F_\phi =-\left( \frac{\partial W}{\partial \phi }\right) ^{*} 
$, $F_\chi =-\left( \frac{\partial W}{\partial \chi }\right) ^{*}$.

\subsection{Superpotential and Scalar Potential}

Let us consider a superpotential, which written in terms of the scalar
fields $\phi $ and $\chi $, is given by 
\begin{equation}
W=\lambda (\phi ^2-a^2)\chi +\frac 13\mu \chi ^3.  \label{e9}
\end{equation}

\noindent We then have the Yukawa coupling terms $Y_{\phi \phi }=2\lambda
\chi $, $Y_{\chi \chi }=2\mu \chi $, $Y_{\phi \chi }=2\lambda \phi $, and
the auxiliary fields are given by $-F_\phi ^{*}=2\lambda \phi \chi $, $%
-F_\chi ^{*}=\lambda (\phi ^2-a^2)+\mu \chi ^2$. The Yukawa part of the
Lagrangian can then be written out as 
\begin{equation}
\begin{array}{ll}
L_Y= & -[\lambda \chi \psi _\phi \psi _\phi +\mu \chi \psi _\chi \psi _\chi
+2\lambda \phi \psi _\phi \psi _\chi ] \\ 
& -[\lambda \chi ^{*}\bar{\psi}_\phi \bar{\psi}_\phi +\mu \chi ^{*}\bar{\psi}%
_\chi \bar{\psi}_\chi +2\lambda \phi ^{*}\bar{\psi}_\phi \bar{\psi}_\chi ].
\end{array}
\label{e10}
\end{equation}

\noindent The scalar potential is given by 
\begin{equation}
V=4\lambda ^2|\phi \chi |^2+|\lambda (\phi ^2-a^2)+\mu \chi ^2|^2.
\label{e11}
\end{equation}

\noindent Note that $V\geq 0$, and vacuum states for which $V=0$ are
supersymmetric vacuum states.

\subsection{Vacuum States}

The potential is $V=F_\phi ^{*}F_\phi +F_\chi ^{*}F_\chi \geq 0$ and the
vacuum states are located by 
\begin{equation}
\begin{array}{ll}
\frac{\partial V}{\partial \phi ^{*}}=F_\phi ^{*}\frac{\partial F_\phi }{%
\partial \phi ^{*}}+F_\chi ^{*}\frac{\partial F_\chi }{\partial \phi ^{*}} & 
=4\lambda ^2|\chi |^2\phi +2\lambda \phi ^{*}[\lambda (\phi ^2-a^2)+\mu \chi
^2]=0, \\ 
\frac{\partial V}{\partial \chi ^{*}}=F_\phi ^{*}\frac{\partial F_\phi }{%
\partial \chi ^{*}}+F_\chi ^{*}\frac{\partial F_\chi }{\partial \chi ^{*}} & 
=4\lambda ^2|\phi |^2\chi +2\mu \chi ^{*}[\lambda (\phi ^2-a^2)+\mu \chi
^2]=0.
\end{array}
\label{e12}
\end{equation}

\noindent Supersymmetric vacuum states are solutions of $V=0$, which is
equivalent to the conditions $F_\phi =0$, $F_\chi =0$. Using $-F_\phi
^{*}=2\lambda \phi \chi $ and $-F_\chi ^{*}=\lambda (\phi ^2-a^2)+\mu \chi
^2 $, we see that there are two possible sets of vacuum states: (1) $\phi
=\pm a $, $\chi =0$, and (2) $\phi =0$, $\chi =\pm \sqrt{\frac \lambda \mu }%
a\equiv \pm \chi _0$. These two sets of vacuum states are energetically
degenerate and supersymmetric ($V=0$). We will focus our attention upon the
first set of vacuum states where $\phi =\pm a$, $\chi =0$. A broken $Z_2$
symmetry associated with $\phi $ gives rise to a domain wall, and a discrete 
$Z_2$ symmetry associated with $\chi $ gets broken in the core of the wall,
giving rise to a $\chi $ condensate and domain ribbons inside the wall.

\section{Domain Wall and Domain Ribbons}

\subsection{Domain Wall}

Let us now focus on the real bosonic sector of the model, where $\psi _\phi
=\psi _\chi =0$ and $%
%TCIMACRO{\func{Im}}
%BeginExpansion
\mathop{\rm Im}
%EndExpansion
(\phi )=%
%TCIMACRO{\func{Im}}
%BeginExpansion
\mathop{\rm Im}
%EndExpansion
(\chi )=0$, i.e. the scalar fields $\phi $ and $\chi $ are real-valued in
this sector. Then the field equations for the scalars $\phi $ and $\chi $ in
the real bosonic sector, obtained, e.g., from $\Box \phi +\left( \frac{%
\partial V}{\partial \phi ^{*}}\right) |_{%
%TCIMACRO{\func{Im} }
%BeginExpansion
\mathop{\rm Im}
%EndExpansion
(\phi )=%
%TCIMACRO{\func{Im} }
%BeginExpansion
\mathop{\rm Im}
%EndExpansion
(\chi )=0}=0$, etc. are given by 
\begin{equation}
\Box \phi +4\lambda ^2\chi ^2\phi +2\lambda \phi [\lambda (\phi ^2-a^2)+\mu
\chi ^2]=0,  \label{e13}
\end{equation}
\begin{equation}
\Box \chi +4\lambda ^2\phi ^2\chi +2\mu \chi [\lambda (\phi ^2-a^2)+\mu \chi
^2]=0.  \label{e14}
\end{equation}

\noindent where $\Box =\partial _0^2-\nabla ^2$. If we assume that the
vacuum states which are realized are given by $\phi =\pm a$, $\chi =0$,
then, when $\chi $ is set equal to zero, a domain wall solution is admitted
for the field $\phi $, with $\phi $ interpolating between the asymptotic
values $\phi =\pm a$. The domain wall solution, describing a wall centered
on the $x-y$ plane ($z=0$), is of the form $\phi (z)=a\tanh \frac z\Delta $,
where $\Delta $ represents the thickness of the wall. It will often be
convenient to approximate the domain wall by a slab of thickness $\Delta $
inside of which $\phi =0$, with $\phi =\pm a$ outside.

We now follow the line of reasoning used by Witten\cite{witten} to examine
the formation of a scalar condensate inside a superconducting cosmic string.
We see that an examination of the field $\chi $ inside the domain wall,
where we take $\phi =0$, indicates that, for a certain parameter range, the
minimal energy configuration of $\chi $ is not given by $\chi =0$, but
rather by $\chi =\pm \chi _0$, where $\chi _0=\sqrt{\frac \lambda \mu }a$.
In this case there are two energetically degenerate ground states given by $%
\chi =\pm \chi _0$ within the core of the wall. Taking the gradient energy
of the field $\chi $ into account, it can be seen that there is a range of
parameters for which $\chi =0$ is, in fact, an unstable solution inside the
wall. This follows by considering small fluctuations of $\chi $ about the
value $\chi =0$. Writing $\chi =F(z)\sin (\omega t)$, and applying this to
the equation of motion for $\chi $ gives 
\begin{equation}
-\partial _z^2F+(2\mu \lambda +4\lambda ^2)[a^2\tanh ^2(\frac z\Delta
)]F=(\omega ^2+2\mu \lambda a^2)F\equiv EF,\,\,\,\,\,E=(\omega ^2+2\mu
\lambda a^2).  \label{e15}
\end{equation}

\noindent Then, for a normalizable bound state for which $E<2\lambda \mu a^2$%
, we have $\omega ^2<0$. We can therefore conclude that there is a parameter
range for which the solution $\chi =0$ is unstable inside the domain wall,
and a scalar condensate with $\chi =\pm \chi _0$ tends to form in the core
of the wall. It will be assumed that the model parameters do indeed occupy a
range for which the condensate formation is energetically favorable.

\subsection{Domain Ribbons}

When the $\chi $ condensate forms within the domain wall, the field $\chi $
can settle into either a $+\chi _0$ state or a $-\chi _0$ state, since these
two states are energetically degenerate. One can expect that domains of
these different states form, but the domains will be uncorrelated beyond
some coherence length $\xi $; i.e., we expect there to be domains where $%
\chi =+\chi _0$ and domains where $\chi =-\chi _0$. Two different domains
are separated by a region where $\chi =0$, locating the core of a domain
ribbon. The domain ribbon is just a portion of a domain wall within the host
domain wall, with the static domain ribbon (R) behaving like $\chi (x)_R\sim
\pm \chi _0\tanh \frac x{w_R}$ and the antiribbon (\={R}) function behaving
like $\chi (x)_{\bar{R}}=-\chi _R(x),$ where $w_R$ is the thickness of the
ribbon or antiribbon. Domain ribbons form between $\pm \chi _0$ domains and
can be in the form of infinite ribbons or in the form of closed ribbon
loops. A ribbon loop encloses a $\pm \chi _0$ domain and is surrounded by a $%
\mp \chi _0$ domain. Self intersecting loops can fission into smaller loops,
with $\chi $ particle radiation being emitted from the annihilating ribbon
sections. Two different loops can also fuse together to form a larger loop,
with the emission of $\chi $ particles. Ribbon loops can also be formed at
the intersections of an infinite ribbon and an antiribbon. Oscillating
ribbon loops with self intersecting trajectories are expected to decay
mainly through $\chi $ particle production, with a negligible fraction of
the released energy in the form of gravitational radiation. (Further details
can be found in ref.\cite{morrisrib})

\subsection{Fermions}

Let us now look at the response of the fermions to the real scalar field
background. Inside the domain wall (but outside of a ribbon or antiribbon),
taking $\phi =0$ and $\chi =+\chi _0$, (\ref{e10}) becomes 
\begin{equation}
L_Y=-\chi _0[\lambda (\psi _\phi \psi _\phi +\bar{\psi}_\phi \bar{\psi}_\phi
)+\mu (\psi _\chi \psi _\chi +\bar{\psi}_\chi \bar{\psi}_\chi ).  \label{e16}
\end{equation}

\noindent In terms of the Majorana spinors $\Psi _{\phi ,\chi }$ this can be
written as 
\begin{equation}
L_Y=i\chi _0[\lambda \bar{\Psi}_\phi \Psi _\phi +\mu \bar{\Psi}_\chi \Psi
_\chi ].  \label{e17}
\end{equation}

\noindent The Majorana mass term is of the form $L_{mass}=\frac 12iM\bar{\Psi%
}\Psi =-\frac 12M(\psi \psi +\bar{\psi}\bar{\psi})$ for a Majorana fermion
of mass $M$. Therefore, we see that in the domain wall (but outside of a
ribbon or antiribbon) the $\Psi _\phi $ fermion mass is $M_\phi =$ $2\lambda 
$\thinspace $\chi _0$ and the $\Psi _\chi $ fermion mass is $M_\chi =$ $2\mu
\chi _0$.

Also, there is [see (\ref{e10})] a Dirac fermion in the vacuum state where $%
\chi =0$, $\phi =+a$, made from the Weyl spinors $\psi _\phi $, $\psi _\chi $%
: $-2\lambda a[\psi _\phi \psi _\chi +\bar{\psi}_\phi \bar{\psi}_\chi
]=i2\lambda a\bar{\psi}^{\prime }\psi ^{\prime }$, where the Dirac spinor $%
\psi ^{\prime }$ is given in terms of the Weyl two-spinors as $\psi ^{\prime
}=\left( 
\begin{array}{l}
\psi _{\phi \,\alpha } \\ 
\bar{\psi}_\chi ^{\dot{\alpha}}\,
\end{array}
\right) $. (Note that in going from a domain where $\phi =+a$ to one where $%
\phi =-a$, the spinor mass eigenstates change, i.e. the Weyl 2 spinors
undergo a phase rotation $\psi _\phi \rightarrow i\psi _\phi $, etc. and the
Majorana 4 spinors undergo a $\gamma _5$ ``rotation'', $\Psi _\phi
\rightarrow \gamma _5\Psi _\phi $, etc.) The Dirac spinor in the vacuum has
a mass $M^{\prime }=2\lambda a$. So, we have a Dirac fermion in the vacuum
with mass $M^{\prime }=2\lambda a$, and Majorana fermions in the domain
wall, which have masses $M_\phi =2\lambda \chi _0$, $M_\chi =2\mu \chi _0$
outside of a domain ribbon, but become massless inside the core of a domain
ribbon.

Now note [see (\ref{e10}) and (\ref{e17})] that in the core of a domain
ribbon, where $\chi \rightarrow 0$, the Majorana fermions become massless: $%
M_{\phi ,\chi }\rightarrow 0$. We can suspect that there are fermion zero
modes\cite{jackiw} within the domain ribbons.

The situation and the particle masses can be briefly summarized in the
following way. There are three different regions where we can look at field
expectation values (DW=domain wall, DR=domain ribbon):

\smallskip 

\begin{description}
\begin{itemize}
\begin{description}
\item  (I) {\it In vacuum:} $|\phi |=a$, $\chi =0$

\item  (II) {\it Inside DW, outside DR: }$\phi =0$, $|\chi |=\chi _0$, $\chi
_0=\sqrt{\frac \lambda \mu }a$

\item  (III) {\it Inside DR:} $\phi =0$, $\chi =0$.
\end{description}
\end{itemize}
\end{description}

\smallskip\ 

The boson particle masses can be examined from the potential $V$ given by (%
\ref{e11}): 
\begin{equation}
\begin{array}{ll}
m_\phi ^2\equiv \frac{\partial ^2V}{\partial \phi \partial \phi ^{*}} & 
=(2\lambda )^2\left[ |\chi |^2+|\phi |^2\right] , \\ 
m_\chi ^2\equiv \frac{\partial ^2V}{\partial \chi \partial \chi ^{*}} & 
=(2\lambda )^2|\phi |^2+(2\mu )^2|\chi |^2.
\end{array}
\label{e17a}
\end{equation}

The fermion masses come from $L_Y$, given by (\ref{e10}). For Dirac and
Majorana fermions the mass terms are of the form 
\begin{equation}
\begin{array}{ll}
L_{mass}=-m(\psi _1\psi _2+\bar{\psi}_1\bar{\psi}_2)=im(\bar{\Psi}\Psi ), & 
\,\,\,\,\,\Psi =\left( 
\begin{array}{l}
\psi _1 \\ 
\bar{\psi}_2
\end{array}
\right) \,\,\,\,\,\text{(Dirac),} \\ 
L_{mass}=-\frac 12m(\alpha \alpha +\bar{\alpha}\bar{\alpha})=\frac 12im(\bar{%
M}M), & \,\,\,\,\,M=\left( 
\begin{array}{l}
\alpha \\ 
\bar{\alpha}
\end{array}
\right) \,\,\,\,\,\text{(Majorana),}
\end{array}
\label{e17b}
\end{equation}

\noindent and these fermion masses have been looked at previously.

A summary of particle masses in the various regions is given below: 
\[
\begin{array}{ll}
\text{Region I :} & m_\phi =m_\chi =M^{\prime }=2\lambda a \\ 
\text{Region II:} & m_\phi =M_\phi =2\lambda \chi _0,\,\,\,m_\chi =M_\chi
=2\mu \chi _0 \\ 
\text{Region III:} & m_\phi =m_\chi =M_\phi =M_\chi =0
\end{array}
\]

\section{Fermion Zero Modes Inside a Domain Ribbon}

\subsection{Reaction of Fermion Fields to Wall and Ribbon Backgrounds}

Now let's consider the effect of the real scalar fields upon the dynamical
spinor fields by again looking at the spinors in the background fields
described by the domain wall and domain ribbon solutions. For approximation
purposes, we neglect field gradients in the wall and ribbons and we take $%
\phi =0$ inside the domain wall (and inside a ribbon) and $\chi (x)=\chi
_0\tanh \frac x{w_R}$ for the (static) ribbon. The Majorana fields $\Psi
_{\phi ,\chi }$ are given in terms of the Weyl 2-spinors by (\ref{e2}) and
the Lagrangian is given by $L=L_K^B+L_K^F+L_Y-V$. Consider the field
equations for the Majorana spinors inside the domain wall, where $\phi =0$.
These field equations follow from $\frac{\partial L}{\partial \bar{\Psi}}=%
\frac{\partial L_K^F}{\partial \bar{\Psi}}+\frac{\partial L_Y}{\partial \bar{%
\Psi}}=0$, where 
\begin{equation}
L_K^F=\frac i2\bar{\Psi}_\phi \gamma ^\mu \partial _\mu \Psi _\phi +\frac i2%
\bar{\Psi}_\chi \gamma ^\mu \partial _\mu \Psi _\chi \,\,,  \label{e19}
\end{equation}
\begin{equation}
\begin{array}{ll}
L_Y & =-[\lambda \chi (\psi _\phi \psi _\phi +\bar{\psi}_\phi \bar{\psi}%
_\phi )+\mu \chi (\psi _\chi \psi _\chi +\bar{\psi}_\chi \bar{\psi}_\chi )
\\ 
& =i\left[ \lambda \chi \bar{\Psi}_\phi \Psi _\phi +\mu \chi \bar{\Psi}_\chi
\Psi _\chi \right] .
\end{array}
\label{e20}
\end{equation}

\noindent Therefore, the field equations for the Majorana fields, in the
background of the domain wall and ribbon fields only [i.e. inside the domain
wall where we assume that $\phi =0$ and $\chi ^{*}=\chi $ ] are given by 
\begin{equation}
\begin{array}{l}
\gamma ^\mu \partial _\mu \Psi _\phi +2\lambda \chi \Psi _\phi =0, \\ 
\gamma ^\mu \partial _\mu \Psi _\chi +2\mu \chi \Psi _\chi =0.
\end{array}
\label{e21}
\end{equation}

\noindent [These equations do not include any descriptions of the possible
interactions of the fermions with scalar field excitations (where, e.g., we
could more generally write $\phi =\phi _{wall}+\delta \phi $, $\chi =\chi
_{ribbon}+\delta \chi $ ).] Again, from the field equations we see that
inside the wall, but outside a ribbon ($\phi =0$, $\chi =+\chi _0$), the
Majorana fermion masses are $M_\phi =2\lambda \chi _0$, $M_\chi =2\mu \chi
_0 $, and inside a ribbon ($\phi =0$, $\chi \rightarrow 0$) the Majorana
fermions become massless, $M_{\phi ,\chi }\rightarrow 0$.

\subsection{Static Zero Modes}

To search for static Majorana zero modes inside a ribbon, let's assume $\Psi
_{\phi ,\chi }=\Psi _{\phi ,\chi }(x)$, and use the fact that $(\gamma
^1)^2=1$, along with $\chi (x)=\chi _0\tanh \frac x{w_R}$ for the
description of a ribbon. Furthermore, since the equations for $\Psi _\phi $
and $\Psi _\chi $, given by (\ref{e21}) are similar and decoupled, let's
only deal with the equation for $\Psi _\phi $ and drop the subscript $\phi $
for now; i.e. $\Psi _\phi \rightarrow \Psi $. The equation for $\Psi $
therefore becomes 
\begin{equation}
\gamma ^1\partial _x\Psi (x)+2\lambda \chi (x)\Psi (x)=0,  \label{e22}
\end{equation}

\noindent where $\gamma ^1=i\left( 
\begin{array}{ll}
0 & \sigma _1 \\ 
-\sigma _1 & 0
\end{array}
\right) $, with $\left\{ \gamma ^\mu ,\gamma ^\nu \right\} =-2g^{\mu \nu }$,
and $(\gamma ^1)^2=1$.

Multiplying (\ref{e22}) by $\gamma ^1$ gives

\begin{equation}
\partial _x\Psi =-2\lambda \chi \gamma ^1\Psi .  \label{e22a}
\end{equation}

\noindent Let us now write the Majorana 4-spinor $\Psi $ in terms of
2-spinors $\eta $ and $\xi $: $\Psi =\left( 
\begin{array}{l}
\eta \\ 
\xi
\end{array}
\right) $. We then have $\gamma ^1\Psi =i\left( 
\begin{array}{ll}
0 & \sigma _1 \\ 
-\sigma _1 & 0
\end{array}
\right) \left( 
\begin{array}{l}
\eta \\ 
\xi
\end{array}
\right) =i\left( 
\begin{array}{l}
\sigma _1\xi \\ 
-\sigma _1\eta
\end{array}
\right) $. Therefore, 
\begin{equation}
\partial _x\left( 
\begin{array}{l}
\eta \\ 
\xi
\end{array}
\right) =-2i\lambda \chi \left( 
\begin{array}{l}
\sigma _1\xi \\ 
-\sigma _1\eta
\end{array}
\right) ,\,\,\,\,\,\,\,\,\,\,\sigma _1=\left( 
\begin{array}{ll}
0 & 1 \\ 
1 & 0
\end{array}
\right) ,\,\,\,\,\,\,\,(\sigma _1)^2=1.  \label{e22b}
\end{equation}

The equations for $\eta $ and $\xi $ can be decoupled by writing 
\begin{equation}
\xi =-i\sigma _1\eta ,\,\,\,\,\,\,\,\,\,\,\eta =i\sigma _1\xi .  \label{e23}
\end{equation}

\noindent Then, by (\ref{e22b}) and (\ref{e23}), 
\begin{equation}
\partial _x\eta =-2\lambda \chi \eta ,\,\,\,\,\,\,\,\,\,\,\partial _x\xi
=-2\lambda \chi \xi ,\,\,\,\,\,\,\,\,\,\,\Psi =\left( 
\begin{array}{l}
\eta \\ 
\xi
\end{array}
\right) =\left( 
\begin{array}{l}
\eta \\ 
-i\sigma _1\eta
\end{array}
\right) .  \label{e24}
\end{equation}

\noindent A solution is given by 
\begin{equation}
\eta =\tau \exp \left[ -2\lambda \int_0^x\chi (x^{\prime })dx^{\prime
}\right] =\tau \left[ \cosh \frac x{w_R}\right] ^{-2},  \label{e25}
\end{equation}

\noindent where $\tau $ is an arbitrary constant Weyl 2-spinor and where $%
w_R=\frac 1{\lambda \chi _0}$ .

The Majorana condition $\Psi _C=-\gamma ^2\Psi ^{*}=\Psi $, (where $\Psi _C$
is the charge conjugate of $\Psi )$ i.e. 
\begin{equation}
\Psi =\left( 
\begin{array}{l}
\eta \\ 
\xi
\end{array}
\right) =\left( 
\begin{array}{l}
\eta \\ 
i\sigma _2\eta ^{*}
\end{array}
\right) ,  \label{e26}
\end{equation}

\noindent must also be satisfied. Upon comparing (\ref{e24}) and (\ref{e26}%
), we have $\sigma _2\eta ^{*}=-\sigma _1\eta $, or $\sigma _1\sigma _2\eta
^{*}=-\eta $, so that with the help of $\sigma _1\sigma _2=i\sigma _3$, we
get $\eta ^{*}=i\sigma _3\eta $. We must therefore require that $\tau
^{*}=i\sigma _3\tau $. We therefore have for our present case the static
zero mode solutions 
\begin{equation}
\begin{array}{ll}
\Psi _\phi =\left( 
\begin{array}{l}
\eta \\ 
\xi
\end{array}
\right) , & \eta =\tau \exp \left[ -2\lambda \int_0^x\chi (x^{\prime
})dx^{\prime }\right] , \\ 
\Psi _\chi =\left( 
\begin{array}{l}
\eta ^{\prime } \\ 
\xi ^{\prime }
\end{array}
\right) , & \eta ^{\prime }=\tau ^{\prime }\exp \left[ -2\mu \int_0^x\chi
(x^{\prime })dx^{\prime }\right] ,
\end{array}
\label{e26a}
\end{equation}

\noindent where $\xi =-i\sigma _1\eta $, $\xi ^{\prime }=-i\sigma _1\eta
^{\prime }$. These solutions describe static Majorana zero modes localized
within the domain ribbon.

\subsection{Traveling Waves}

Let us now regard $\Psi $ to be a function of $x$, $y$, and $t$, i.e., $\Psi
(x,y,t)=\left( 
\begin{array}{l}
\eta (x,y,t) \\ 
-i\sigma _1\eta (x,y,t)
\end{array}
\right) $, where $\eta (x,y,t)=\tau (y,t)\left[ \cosh \frac xw\right] ^{-2}$%
. Then (\ref{e21}) implies that 
\begin{equation}
(\gamma ^0\partial _0+\gamma _2\partial _2)\left( 
\begin{array}{l}
\tau (y,t) \\ 
-i\sigma _1\tau (y,t)
\end{array}
\right) \left[ \cosh \frac xw\right] ^{-2}=0,  \label{e27}
\end{equation}

\noindent which is solved by 
\begin{equation}
(\partial _0+\sigma _2\partial _2)\tau (y,t)=0.  \label{e28}
\end{equation}

\noindent This can be seen by multiplying (\ref{e27}) by $\gamma ^1$ and
using $\gamma ^0\gamma ^2=\left( 
\begin{array}{ll}
\sigma _2 & 0 \\ 
0 & -\sigma _2
\end{array}
\right) $, so that (\ref{e27}) reduces to the set of equations $(\partial
_0+\sigma _2\partial _2)\tau =0,$ and $(\partial _0-\sigma _2\partial
_2)\sigma _1\tau =0$, and the second equation is automatically solved when
the first equation is solved. Then writing $\tau =\left( 
\begin{array}{c}
\tau _1 \\ 
\tau _2
\end{array}
\right) $ (\ref{e28}) can be written explicitly as 
\begin{equation}
\partial _0\tau _1-i\partial _2\tau _2=0,\,\,\,\,\,\,\,\partial _0\tau
_2+i\partial _2\tau _1=0.  \label{e29}
\end{equation}

\noindent These can be combined to give 
\begin{equation}
(\partial _0^2-\partial _2^2)\tau _{1,2}=0,  \label{e29a}
\end{equation}

\noindent which is solved by $\tau _{1,2}(y,t)=\tau _{1,2}(y\pm t)$.
Therefore, $\tau $ can be written as 
\begin{equation}
\tau (y,t)=\left( 
\begin{array}{l}
\tau _1(y\pm t) \\ 
\tau _2(y\pm t)
\end{array}
\right) .  \label{e31}
\end{equation}

\noindent so that $\tau $, and hence $\Psi $, can contain a linear
combination of ``up'' and ``down'' moving waves.

\section{Fermi Gas in Domain Ribbon Loops}

\subsection{One-Dimensional Fermi Gas}

Let's consider the case where fermions occupy the interior of a domain
ribbon, so that in the singular, thin ribbon limit, there is a
one-dimensional Fermi gas. (The two Majorana spinors in the domain ribbon
can be related to a Dirac spinor.) The number of states of spin 1/2 fermions
with momentum between $p_x$ and $p_x+dp_x$ in a length $L$ is 
\begin{equation}
\rho (p_x)dp_x=\frac{gL}{2\pi \hbar }dp_x,  \label{ae5}
\end{equation}

\noindent where $g=2$ is the number of spin states for a spin 1/2 fermion.
In the ground state there is a momentum spread from $p_x=-p_F$ to $p_x=p_F$,
so that the number of fermions in the ground state is 
\begin{equation}
N=\int_{-p_F}^{p_F}\rho (p_x)dp_x=\frac{gL}{\pi \hbar }\,p_F  \label{ae6}
\end{equation}

\noindent which implies that 
\begin{equation}
p_F=\frac{\pi \hbar N}{gL}.  \label{ae7}
\end{equation}

\noindent The total energy of fermions in the ground state is 
\begin{equation}
E_F=\int_{-p_F}^{p_F}\rho (p_x)\epsilon (p_x)dp_x,  \label{ae8}
\end{equation}

For the case of massless fermions, $\epsilon =pc=|p_x|c$, and we therefore
get 
\begin{equation}
E_F=\frac{\pi \hbar cN^2}{2gL}=\frac{\pi \hbar cN^2}{4L},  \label{ae9}
\end{equation}

\noindent where we have set $g=2$ for one species of spin $1/2$ fermion.

\subsection{Massless Fermi Gas in a Ribbon Loop}

Now consider a ribbon loop of length $L$ to be inhabited by a
one-dimensional Fermi gas of massless fermions, with a ``ribbon field'' mass
function $\chi (x)$, which vanishes in the ribbon core at $x=0$. (We now set 
$\hbar =c=1$.) The energy per unit length of ribbon is $\mu _R$ and the
energy of the Fermi gas is $E_F$. The total energy for the ribbon loop is
therefore 
\begin{equation}
E=\mu _RL+E_F=\mu _RL+\frac{\pi N^2}{2gL},  \label{ae11}
\end{equation}

\noindent where $N$ is the total fermion number for the fermions inhabiting
the loop. For a fixed value of $N$, the energy is minimized for 
\begin{equation}
L=\left( \frac \pi {2g\mu _R}\right) ^{1/2}N  \label{ae12}
\end{equation}

\noindent which, for a circular ribbon loop with $L=2\pi R$, would
correspond to a radius of $R=\left( \pi \mu _R\right) ^{-1/2}\frac N4$.

\noindent For a loop of length $L=\sqrt{\pi /\mu _R}(N/2)$, the total energy
of the loop is 
\begin{equation}
E=\sqrt{\pi \mu _R}N.  \label{ae13}
\end{equation}

However, we can notice that the loop is evidently unstable against
flattening, since the configuration energy $E$ depends upon the loop length $%
L$, but is independent of the loop area, which could be decreased while
keeping the length $L$ constant. Therefore the loop may flatten (or have
self intersecting trajectories) and subsequently fragment into smaller
loops. This process may be continued resulting in the production of many
smaller loops, but we expect this process to halt when the thin ribbon
approximation breaks down, and the solitonic structure of the ribbon becomes
important. (A similar type of reasoning has been used previously by
MacPherson and Campbell\cite{mc} in the description of the collapse of three
dimensional false vacuum bags to form ``Fermi balls''.) Let us assume that
stable circular loops of radius $R$ are produced at the end of this
fragmentation process. We take the minimal loop radius, where the thin
ribbon wall approximation begins to break down, to be roughly $R_{\min }\sim
\nu w_R$, where $w_R$ is the ribbon width, or thickness, and a reasonable
guess for $\nu $ may be roughly 1-10.

A static, straight domain ribbon has a profile given by $\chi _R=\chi
_0\tanh (x/w_R)$, where $w_R=1/(\lambda \chi _0)$ is the thickness of the
ribbon, and we estimate the energy density of this configuration\cite
{morrisrib} to be $T_{00}^{(R)}\sim (\partial _x\chi )^2=\lambda ^2\chi _0^4$%
sech$^4(x/w_R)$. The energy per unit area of the ribbon is then roughly $%
\Sigma \sim (\chi _0^2/w_R)w_R=\lambda \chi _0^3$. We multiply this by the
thickness $\Delta =1/(\lambda a)$ of the domain wall to get an estimate of
the energy per unit length, $\mu _R$,  of the ribbon:
\begin{equation}
\mu _R\sim \Sigma \Delta \sim \frac{\chi _0^3}a=\left( \frac \lambda \mu
\right) ^{3/2}a^2.  \label{ae14}
\end{equation}

\noindent By (\ref{ae12}) $L\sim N/\sqrt{\mu _R}$ so that upon setting $%
L/2\pi $ equal to $R_{\min }\sim \nu /(\lambda \chi _0)$, we get 
\begin{equation}
N\sim \frac{2\pi \nu }\lambda \sqrt{\frac{\chi _0}a}=2\pi \nu \left( \frac 1{%
\lambda ^3\mu }\right) ^{1/4}  \label{ae14a}
\end{equation}

\noindent as an estimate for the number of fermions that occupy a stabilized
ribbon loop.

By (\ref{ae13}) the mass of a stabilized loop is roughly $E\sim N\sqrt{\mu _R%
}$, which by (\ref{ae14}) and (\ref{ae14a}), gives
\begin{equation}
E\sim \frac{2\pi \nu }{\lambda a}\chi _0^2=\frac{2\pi \nu }\mu a.
\label{ae14b}
\end{equation}

\noindent At the GUT scale, the mass of the ribbon loop is roughly $E\sim
(2\pi \nu /\mu )10^{16}$ GeV, while at the electroweak scale, $E\sim (2\pi
\nu /\mu )10^3$ GeV.

\section{Summary}

Topological defects represent interesting nonperturbative field theoretic
solutions, but they are also interesting because they may have been
physically realized in the early Universe during symmetry breaking phase
transitions. This defect production may have taken place when the Universe
existed in a supersymmetric phase, and it is therefore of interest to
investigate defects within a supersymmetric context. Realistic
supersymmetric theories contain interacting chiral superfields, so we have
examined an example of a type of supersymmetric model where interactions can
yield defects with a nontrivial internal topological structure. This extends
some previous work on supersymmetric defects and structured nonsupersymmeric
defects. Supersymmetry also couples fermions to the scalar fields, so that
there may be fermionic effects introduced, such as the existence of zero
modes and degeneracy pressure in defects.

We have considered a model admitting the simplest type of topological
defect, a domain wall. The simple superpotential allows interactions between
two scalar fields, with the result that a real scalar condensate can form
within the wall. There is a distribution of $\pm \chi _0$ condensate domains
within the wall, and at the interface between two different domains there
must exist a topological ``domain ribbon''. The ribbon can support fermion
zero modes, which have been described analytically. In general, there will
be fermionic excitations above the zero mode, describing fermionic particles
trapped within the ribbon. These fermions give rise to a Fermi gas pressure,
which can help to stabilize ribbon loops so that they do not completely
disappear because of fissioning due to self intersecting loop trajectories,
but perhaps stabilize in the form of microscopic particle sized loops.
However, a complete description of the dynamics of the infinite ribbons and
the multiple loops occupying the domain wall may be complicated, with
fission and fusion processes taking place due to ribbon and loop
interactions. It could be the case that essentially all of the loops in a
domain wall eventually annihilate one another away, which could leave a
domain wall populated with fermions, depending upon the relative values of
the model parameters. At any rate, it can be seen that interactions in
supersymmetric field theories of topological defects can give rise to
bosonic and fermionic effects that may often be otherwise ignored or
overlooked.

\smallskip\ 

{\bf Acknowledgement}

\smallskip\ 

I wish to thank D. Bazeia for discussions related to this work.

\appendix 

\section{Conventions}

Some of the notations and conventions are briefly listed here. A metric $%
g_{\mu \nu }$ is used with signature $(+,-,-,-)$. Aside from the metric, the
notation, conventions, and gamma matrices used conform to those of ref.\cite
{srivastava} The gamma matrices can be written in the form 
\begin{equation}  \label{a1}
\gamma ^\mu =i\left( 
\begin{array}{cc}
0 & \sigma ^\mu \\ 
\bar \sigma ^\mu & 0
\end{array}
\right)
\end{equation}

\noindent with 
\begin{equation}  \label{a2}
\sigma ^\mu =(1,{\bf \vec \sigma })\,\,,\,\,\,\,\,\,\,\,\,\,\bar \sigma ^\mu
=(1,-{\bf \vec \sigma })\,\,,
\end{equation}

\noindent where ${\bf \vec \sigma }$ represents the Pauli matrices. Then 
\begin{equation}  \label{a3}
\gamma ^0=i\left( 
\begin{array}{cc}
0 & 1 \\ 
1 & 0
\end{array}
\right) ,\,\,\,\,\,\,\,\,\,\,\gamma ^k=i\left( 
\begin{array}{cc}
0 & \sigma _k \\ 
-\sigma _k & 0
\end{array}
\right) ,\,\,\,\,\,k=1,2,3,
\end{equation}

\noindent and $\gamma _5$ is given by 
\begin{equation}  \label{a4}
\gamma _5=\gamma ^0\gamma ^1\gamma ^2\gamma ^3=i\left( 
\begin{array}{cc}
1 & 0 \\ 
0 & -1
\end{array}
\right) .
\end{equation}

\noindent The gamma matrices have the properties 
\begin{equation}  \label{a5}
\{\gamma ^\mu ,\gamma ^\nu \}=-2g^{\mu \nu },\,\,\,\,\{\gamma ^\mu ,\gamma
_5\}=0,\,\,\,\,\gamma _5^{\dagger }=-\gamma _5,\,\,\,\,(\gamma _5)^2=-1.
\end{equation}

\noindent A Majorana 4-spinor $\Psi $ is expressed in terms of the Weyl
2-spinors $\psi $ and $\bar \psi $ by $\Psi =\left( 
\begin{array}{c}
\psi _\alpha \\ 
\bar \psi ^{\dot \alpha }
\end{array}
\right) $ and we use the summation conventions for Weyl spinors [with $\bar 
\psi ^{\dot \alpha }=(\psi ^\alpha )^{*}$] 
\begin{equation}  \label{a6}
\xi \psi \equiv \xi ^\alpha \psi _\alpha ,\,\,\,\,\bar \xi \bar \psi \equiv 
\bar \xi _{\dot \alpha }\bar \psi ^{\dot \alpha },\,\,\,\,\alpha
=1,2,\,\,\,\,\dot \alpha =1,2,
\end{equation}

\noindent with $\varepsilon $ metric tensors (for raising and lowering Weyl
spinor indices) 
\begin{equation}
(\varepsilon ^{\alpha \beta })=(\varepsilon ^{\dot{\alpha}\dot{\beta}%
})=i\sigma _2,\,\,(\varepsilon _{\alpha \beta })=(\varepsilon _{\dot{\alpha}%
\dot{\beta}})=-i\sigma _2,\,\,\,\,\varepsilon ^{12}=1=\varepsilon ^{\dot{1}%
\dot{2}}.  \label{a7}
\end{equation}

\newpage\

\end{document}